\documentclass[11pt]{article}
\usepackage{amsmath} % Better maths support
\usepackage{amssymb}
\usepackage{amsthm}
\usepackage[vcentermath,enableskew]{youngtab}
\usepackage{mathrsfs}
\usepackage{latexsym}
\newcommand{\be}{\begin{equation}}
\newcommand{\ee}{\end{equation}}
\newcommand{\ea}{\end{array}}
\newcommand{\beqa}{\begin{eqnarray}}
\newcommand{\eeqa}{\end{eqnarray}}
\newcommand{\nin}{\noindent}
\newcommand{\nn}{\nonumber}
\newcommand{\Tr}{\mbox{Tr}}
\def\Dots{\cdot\cdot}
\def\CP{{\mathbb C}P}
\def\half{\frac{1}{2}}
\def\D{ D \! \! \! \!  / }
\def\bull{\bullet}
\def\cross{\times}
\begin{document}

%\special{papersize=8.26in,11.69in}
%\textwidth15.0cm
%\textheight22.0cm
%\baselineskip1.0cm
%\setlength{\topmargin}{-1cm}
%\addtolength{\textheight}{1cm}
%\oddsidemargin+1.2cm
%\evensidemargin-1.2cm
%\topmargin=-.3in
%\textheight=8.3in
%\textwidth=7.2in
%\textwidth=6.7in
%------------------------------------------------------
\medskip

\begin{center}
{\Large\bf A projective Dirac operator on $\CP^2$ \\ within fuzzy geometry}

\vspace{3pt}

{\bf \Large }
\vskip1.3cm

{\large I. Huet}\footnote{e-mail address: idrish.huet@uni-jena.de}
\\[1.5ex]

\vspace{0.3cm}

{\it 
Theoretisch-Physikalisches Institut \\
Friedrich Schiller Universit\"at, Jena, \\
Max-Wien-Platz 1, D-07743,\\
Th\"uringen, Germany}
\end{center}
%\centerline{\today}
\vspace{1cm}
\begin{center}
 {\large \bf Abstract:}
\begin{quotation}
\nin We propose an ansatz for the commutative canonical spin$_{c}$ Dirac operator on $\CP^2$ in a global geometric approach using the right invariant (left action-) induced vector fields from $SU(3)$. This ansatz is suitable for noncommutative generalisation within the framework of fuzzy geometry. Along the way we identify the physical spinors and construct the canonical spin$_c$ bundle in this formulation. The chira\-lity operator is also given in two equivalent forms. Finally, using representation theory we obtain the eigenspinors and calculate the full spectrum. We use an argument from the fuzzy complex projective space $\CP^2_F$ based on the fuzzy analogue of the unprojected spin$_c$ bundle to show that our commutative projected spin$_c$ bundle has the correct $SU(3)$-representation content. 
\end{quotation}
\end{center}
%\vfill\eject
\pagestyle{plain}
\setcounter{page}{1}
\setcounter{footnote}{0}

\section{Introduction}

This paper is motivated by the long standing problem in fuzzy quantum field theory of providing 4-dimensional (4D) QED theories in a fuzzy space. Fuzzy spaces are a special kind of 
noncommutative geometries wherein the algebra of functions is approximated by a sequence of finite matrix algebras, they have been subjected to both theoretical and numerical 
intensive analysis in the last decade \cite{Marco}. We are concerned with a formulation of 4D fuzzy QED, along this line important examples so far are the direct product of fuzzy 
spheres $S^2_F \times S^2_F$ treated in \cite{Rodel,Stein}, where a pure gauge field was studied, the case of $S_F^2$, studied in \cite{Klimcik} for self interacting fields and the 
$q-$deformed fuzzy sphere treated in \cite{Harikumar}. A genuine theory on $S^4$ (which could in principle be obtained as an effective theory on an squashed 
$\CP^3$ \cite{Julieta, Baer, Grosse}) or $\CP^2$ is still lacking. A study along this direction was undertaken in \cite{Stein}, where a gauge theory was formulated, in a similar 
spirit to this paper, also on $\CP^2_F$.
Both of these 4-dimensional examples are related to complex projective spaces $\CP^N$, which in their own right are interesting objects, for instance, they are relevant models for 
the higher dimensional quantum Hall effect \cite{Karabali} and it has been suggested that the chiral zero modes of the gauged Dirac operator are connected to the origin of the 
standard model fermion spectrum \cite{Briany}. These spaces provided the first example of an infinite family of fuzzy spaces \cite{CPNF}, which was later generalised to Grassmann 
manifolds and many others \cite{Saemann, Brianz}.

Usually the traditional approach of classical differential geometry, involving local coordinate systems along with coordinate patches, does not make much sense in noncommutative 
geometry because points cannot be located and open sets seem unnatural; coordinates in a background noncommutative space, however, may make sense. This is the case in the archetypical 
fuzzy sphere \cite{Madore} $S^2_F$ built from the usual sphere embedded in $\mathbb{R}^3$ where the usual coordinates $(x,y,z)$ are replaced by the $SU(2)$ generators, $L_i$, in the 
spin $s$ irreducible representation; functions on the fuzzy sphere are just polynomials on the generators taking into account the restriction $L_i L_i = s(s+1){\bf 1}$, which plays 
the role of the embedding equation $x^2+ y^2+z^2 = r^2$, and may be represented as arbitrary $2s+1$ size complex square matrices. The construction of fuzzy complex projective spaces, 
$\CP^N_F$, was originally given by O'Connor et al in \cite{CPNF}, wherein the global coordinate system that describes $S^2$ was generalised to $\CP^N$. In this paper we focus in the 
special case of $\CP^2$ and propose an ansatz for the Dirac operator on this space, which is formulated in the forementioned global coordinate system. The Dirac operator is an object 
of great relevance and subject to intense study within noncommutative geometry; it is often the tool used to define metric- and differential-geometric concepts in it 
\cite{Connes, Padman}.

The Dirac operator on $\CP^N$ has been known for a long time\cite{Cah, Seifarth}, in \cite{Bal2} it was formulated on terms of right action differential operators (universal covariant 
derivatives \cite{Batos}). In a previous work \cite{spinor} a fuzzy analogue was constructed for $\CP^N_F$ using a Schwinger-Fock construction for the right action differential 
operators. Such formulation, however, seems ill suited to couple the fermions with gauge fields \cite{Doc}. An idea to go around this difficulty is to find a formulation of the Dirac 
operator using left action differential operators instead. This approach would require projecting both, the spinor and the gauge fields onto the physically relevant components for 
$\CP^N_F$. In this paper we find an ansatz for the Dirac operator and show how this projection can be realised in the case of commutative $\CP^2$ by defining the physical spin$_c$ 
bundle $\mathfrak{S} \subset S$. We give also the fuzzy analogue of the redundant (unprojected) spin$_c$ bundle $S$, and show that the commutative limits of certain projective modules 
give the correct $SU(3)$ content for the canonical spin$_c$ bundle. This procedure provides an elegant way of finding the representation content of the subbundles and thus to identify 
correctly $\mathfrak{S}$ as the canonical spin$_c$ bundle.

It is known that on $\CP^N$ nonvanishing spinor fields can only be defined when $N$ is odd \cite{Haw, Lawson}, this is expressed by saying that $\CP^{2n+1}$ admits a spin structure. 
On the contrary, when $N$ is even it does not admit a spin structure but rather a similar structure called spin$_c$, in fact there is an infinite number of such structures that can be 
put on $\CP^N$ for arbitrary $N$ \cite{Baer2, Lawson, spinor, Seifarth}, each corresponding to a different spin connection and thus to different Dirac operators. We found that our 
ansatz corresponds to the so called canonical spin$_c$ Dirac operator on $\CP^2$ where the charge of the fermions under global $U(1)$ rotations vanishes; in this case spinors may be 
constructed as $(0,k)-$forms \cite{Witten, spinor}.

The paper is divided as follows: Section 2 presents a quick review of the construction of $\CP^2$ as an $SU(3)$ orbit and introduces some conventions. Section 3 is devoted to explain 
the harmonic decomposition of functions on $\CP^2$, which will later be necessary to find the eigenspinors, and presents some useful structures of the tangent space first introduced 
in \cite{CPNF}. Section 4 presents the ansatz itself and shows consistency with Lichnerowicz's theorem, identifies the spin laplacian, spin covariant derivative and curvature tensor. 
Section 5 gives the construction of the spin$_c$ bundle and the calculation of the spectrum, it also introduces the chirality operator and results regarding the agreement with known 
literature. Section 6 contains our argument based on the fuzzy geometry about the $SU(3)$ content of the relevant subbundles, whose fuzzy analogues appear as projective modules over 
a matrix algebra therein. Appendix A is an auxiliary calculation of the Riemann curvature on reductive coset spaces, appendix B presents the calculation of the spectrum for the 
hypercharge operator on $\CP^N$, which we use to compute the spectrum. Appendix C presents the evaluation of the stability subgroup quadratic Casimir operator in the required 
representations, along with a short proof regarding the choice of representations needed to expand functions on $\CP^2$.

\setcounter{equation}{0}
\bigskip

\section{$\CP^2$ orbit construction review} \label{orbit}

We briefly review the construction presented in \cite{CPNF} of $\CP^N$ as an $SU(N+1)$ orbit, particularising to our case of interest $\CP^2$, and introduce notation along with some definitions.

The Lie algebra $su(3)$ is generated by the set of eight Gell-Man matrices $\lambda_a$, indeed:

\be \label{A00}
\lbrack \lambda_a, \lambda_b \rbrack = 2if^{c}_{~ab} \lambda_c
\ee
The Gell-Man matrices are all traceless, and taken together with the identity form a basis for $3 \times 3$ complex matrices. Their algebra is:

\be \nn
\lambda_a \lambda_b = \frac{2}{3} \delta_{ab}{\bf 1} + ( d^{c}_{~ab} + i f^{c}_{~ab} ) \lambda_c ~~\quad~\quad~~a,b,c=1,\ldots,8.
\ee
In this relation both, the structure constants $f$, and the symmetric traceless $d$ are real $SU(3)$-invariant tensors. From the tracelessness of $\lambda_a$ one sees at once their orthogonality under the trace inner product

$$
\Tr (\lambda_a \lambda_b) = 2 \delta_{ab}.
$$
If we consider the last generator $\lambda_8$ and produce its orbit under the adjoint action of $SU(3)$ in the fundamental representation we obtain $\CP^2$. The stability subgroup for $\lambda_8$ is $S(U(2) \times U(1))$, so we may realise $\CP^2$ as a coset space, namely

\be \nn
  \CP^2 = \{ g \lambda_8 g^{\dagger} : g \in SU(3) \} = SU(3)/ S(U(2)\times U(1)).
\ee
The $\CP^2$ space can be also be reintrepeted as the adjoint orbit of a rank one hermitian fiducial projector given through
\be \label{A0}
\mathcal{P}_0 := \frac{\bf 1}{3} - \frac{\lambda_8}{\sqrt{3}} = \mathcal{P}_0^2 = \mathcal{P}_0^{\dagger}
\ee
resulting in a new projector to which we asign coordinates in the Gell-Man basis:

\be \label{A1}
\mathcal{P} := g \mathcal{P}_0 g^{\dagger} =  \frac{\bf 1}{3} + \frac{\xi^a \lambda_a}{\sqrt{3}}  =\mathcal{P}^{2} = \mathcal{P}^{\dagger}
\ee
Therefore the eight real numbers $\xi^a$ can be seen as a global coordinate system for $\CP^2$; since such system must be redundant these quantities satisfy constraints. Algebraically this is just the requirement that (\ref{A1}) be a projector, the above mentioned constraints are:

\be \label{A2}
 \xi^a \xi_a = 1 , ~~~\quad~~~ d^{c}_{~ab}\xi^a \xi^b = \frac{1}{\sqrt{3}}\xi^c
\ee
and describe the embedding $\CP^2 \hookrightarrow \mathbb{R}^8$. Notice that since the coordinates $\xi^a$ carry the adjoint representation of $SU(3)$ and this is a real self-conjugate representation, we will no longer distinguish between upper and lower indices, so e.g. $\xi^a =\xi_a$, $f^{c}_{~ab}= f_{cab}$, $d^{c}_{~ab}= d_{cab}$, and we keep all as subindices unless otherwise stated. With this notation, for $SU(3)$, the $f$ tensor is totally skew-symmetric and the $d$ tensor is fully symmetric and traceless in each pair of indices.  

We may use relations (\ref{A0}) , (\ref{A1}) and the orthogonality of the Gell-Man basis to show that

\be \label{A3}
\xi_a = -\half \Tr (\lambda_a g \lambda_8 g^{\dagger})
\ee
Setting $g={\bf 1}$ corresponds to what we call henceforth the ``north pole" whose coordinates are ${\xi }^0=(0,\ldots,0,-1)$.

Our main interest is a formulation of the Dirac operator in terms of left action differential operators. For this end we define first the left action of a group element $k \in G$ over $g \in G$ as a map $L_k : G \times G \longrightarrow G$ for any group $G$:

\be \label{A4}
L_{k} : g \mapsto k^{-1}g
\ee
When $k$ is an element infinitesimally close to the identity the generators of the group will induce differential operators as the result of the infinitesimal left action on functions defined over the group, we call these the left action differential operators.
If we consider a special kind of functions, for instance, power series of the coordinates $\xi_a$, we obtain from (\ref{A3}) that the left action differential operators are realised as \cite{CPNF}

\be \label{A5}
\mathcal{L}_a = -i f_{abc}\xi_b \frac{\partial}{\partial \xi_c} 
\ee  
and furnish a representation of the $su(3)$ Lie algebra under commutation:

\be \label{A6}
[\mathcal{L}_a , \mathcal{L}_b] = i f_{abc}\mathcal{L}_c
\ee
We shall give our ansatz for the Dirac operator in terms of these differential operators in section \ref{ansatz}.

\section{Harmonic decomposition and tangent space}

The harmonic decomposition of a square integrable function on a compact Lie group $G$ is given by the Peter-Weyl theorem, this theorem states that such function may be expanded in terms of the matrix elements of all the inequivalent unitary irreducible representations of $G$ (which we label with $J$), call them $D^J_{mn} (g)$:

\be \label{B0}
L^2(G,\mathbb{C}) \ni \varphi(g) = \sum_{J,m,n} \varphi^J_{mn} D^{J}_{mn}(g), \quad \quad \varphi^J_{mn} \in \mathbb{C}
\ee
and that such system of matrix elements is a complete orthogonal set, the inner product defined by an appropriate Haar measure $d\mu (g)$

\be \nn
\int_{G}d\mu (g) (D^J_{mn}(g))^{*}D^K_{pq}(g)  = \delta^{JK}\delta_{mp}\delta_{nq}.
\ee
To consider functions defined on a coset manifold $G/H$ we may consider functions over $G$ and restrict ourselves to a particular class of functions called $H$-equivariant functions, these are functions on $G$ that have the additional property:  

\be \nn
\varphi(g) = \varphi(gh), \quad \quad \quad \mbox{for any} ~g \in G,~~ h \in H
\ee
and are precisely all the genuine functions defined on $G/H$. We are taking a special case, with $G=SU(3)$ a simple, compact and simply connected Lie group.

As pointed out in \cite{Bal2} we can find an harmonic expansion for functions on the coset if we observe that $D^J_{mn}(gh) = D^J_{mq}(g)D^J_{qn}(h)$; for that end we must restrict the sum of (\ref{B0}) only to representations $J$ that when reduced under the subgroup $H$ contain the trivial representation. Also, we must reduce the second subindex of $D$ to an adequate subset $I_0$ for each such representation $J$ so that $D^J_{mi_0}(gh)=D^J_{mi_0}(g)$ when $i_0 \in I_0$. Therefore, a square integrable function on the coset $\varphi \in L^2 (G/H, \mathbb{C})$ may be expanded in harmonic functions in the following manner, choosing any representative $g$:

\be \label{B1}
\varphi(gh) = \varphi(g) = \sum_{\begin{subarray}{1}
J,m \\
i_0 \in I_0 
\end{subarray}} \varphi^J_{m i_0}D^J_{mi_0}(g).
\ee
The left action (\ref{A4}) induces a corresponding left action on the functions $\varphi(g)$ defined as $\tilde{L}_{g_0} : \varphi(g) \mapsto \varphi(g^{-1}_0 g)$. Considering now $g_0$ infinitesimally close to the idenity and left acting on (\ref{B1}):

$$
D^J_{mn}(g_0^{-1}) = \delta_{mn} - i (\epsilon \cdot \mathcal{J}^J)_{mn}
$$
wherein the vector $\epsilon$ is infinitesimally small and $\mathcal{J}^J$ are the (hermitian) generators of $G$ in the $J$ representation.

The left action differential operators (\ref{A5}) are defined by the series expansion for such action on a given  $H$-equivariant function, namely 

\be \label{B2}
\varphi(g^{-1}_0 g) = \varphi(g) + i \epsilon \cdot \mathcal{L}\varphi(g) + \cdots
\ee
Putting these facts together we find that the left action of a generator of the Lie algebra of $G$, labeled by $a$, decomposes into the sum of actions of generators acting on each of the harmonic components of the function, this is:

\be \nn
\mathcal{L}_a \varphi(g) = - \sum_{\begin{subarray}{1}
J,m,n \\ 
i_0 \in I_0
\end{subarray}}\varphi^J_{m i_0} (\mathcal{J}^J_a)_{mn}D^J_{ni_0}(g). 
\ee
It is then straightforward to check that (\ref{A6}) is satisfied by the operators $\mathcal{L}_a$, which are given by (\ref{B2}).
Finally, we would like to tell the details about the harmonic expansion (\ref{B1}) for $\CP^2$, these are given in \cite{CPNF}, \cite{cp2}: A function in $SU(3)$ can be expanded according to (\ref{B0}) as follows

\be \nn
\varphi(g) = \sum_{l_1,~l_2}\sum_{\begin{subarray}{1}
I, I_3,Y \\ 
I', I_3 ', Y'
\end{subarray}} \varphi^{(l_1,l_2)}_{I,I_3,Y;I',I_3 ',Y'} D^{(l_1,l_2)}_{I,I_3,Y; I',I_3 ',Y'}(g).
\ee
In the above the representations of $SU(3)$ are labeled by their Dynkin indices $(l_1,l_2)$ and the basis vectors of a given representation are completely determined by the isospin, third component of isospin and hypercharge $I,I_3,Y$ respectively.

It can be shown (appendix \ref{C}) that only those representations that have $l_1 =l_2$ contain the trivial representation when reduced under $S(U(2) \times U(1))$, call this number $l$, then we label the relevant representations $J$ with $l$. On the other hand the restriction to the index subset $I_0$ corresponds to taking only the column of the matrix that has $I'=I_3 '=Y' =0$, this also ensures that the identity representation appears exactly once for each $l$.

The generalised harmonics on $\CP^2$ are the functions $D^{(l,l)}_{I,I_3,Y;0,0,0}(g):=\mathcal{Y}^l_{I,I_3,Y}(g)=\mathcal{Y}^{l}_{I,I_3,Y}(gh)$, and functions over $\CP^2$ written in such basis will be, from (\ref{B1})

\be \label{B3}
\varphi(g) = \sum_{l,I,I_3,Y} \varphi^{l}_{I,I_3 ,Y} \mathcal{Y}^{l}_{I,I_3,Y}(g)
\ee

The $SU(3)$ representation content of this expansion is given by either

\be \label{B4}
\bigoplus_{l=0}^{\infty} (l,l) \quad \quad \mbox{or} \quad \quad {\bf 1 }\oplus {\bf 8 }\oplus {\bf 27 }\oplus{\bf 64 }\oplus{\bf 125 }\oplus \cdots
\ee

\subsection{Tangent space structure}

The interested reader can consult \cite{CPNF} for details, we will summarize the main resuts that are needed.
Any complex projective plane, in particular $\CP^2$, is a complex manifold; it can then be endowed with a complex structure. A complex structure consists of a type (1,1) real antisymmetric tensor that allows one to decompose, in a globally consistent manner,  the tangent space into holomorphic and antiholomorphic subspaces.

In our approach the complex structure is given, componentwise:

\be \nn
J_{ab} = \frac{2}{\sqrt{3}}f_{abc} \xi_c
\ee
Its square provides us with the $SU(3)$-invariant induced metric on $\CP^2$, which we denote by $P$, through the relation $J^2 = -P$. The components of the metric tensor on $\CP^2$ are, in our coordinate system

\be \nn
P_{ab} = \frac{2}{3}\delta_{ab}+ \frac{2}{\sqrt{3}}d_{abc}\xi_c - \frac{4}{3}\xi_a \xi_b
\ee
The tensor $P$ can be reinterpreted also as a projector onto the tangent space of $\CP^2$, therefore it has rank 4, and the relations

$$ JP=PJ=J= -J^T, \quad \quad J^2 = -P,\quad \quad P^2 = P = P^T, \quad \quad \Tr P = 4$$
hold. 

These two real tensors may be used to define a complex projector onto the (anti-)holomorphic part of the tangent space, usually called by the name of K\"ahler structure:

$$ K_{\pm} = \frac{1}{2}(P \pm i J) $$
A notion of tangentiality may be thus defined using $P$, in this case we say e.g. that the differential operators $\mathcal{L}$ are tangent whilst the coordinates $\xi$ are normal, since $P_{ab}\mathcal{L}_b   = \mathcal{L}_a, ~P_{ab} \xi_b = 0$.

\section{The ansatz} \label{ansatz}

With the Dirac matrices in $\mathbb{R}^8$ and their commutators, 

\be
\{ \gamma_a ,\gamma_b \} = 2 \delta_{ab} {\bf 1}, \quad \quad  \gamma_{ab} := \frac{1}{2} [\gamma_a ,\gamma_b]
\ee
we construct the ansatz for the Dirac operator as 

\be \label{C0}
\D = \gamma_{a} P_{ab} (\mathcal{L}_b+ T_b) = \gamma_a \mathcal{L}_a + \gamma_a P_{ab}T_b :=\gamma_a D_a.
\ee
This ansatz is $SU(3)$ invariant, also notice that we have defined the covariant derivatives $D_a$ and introduced the operators $T_a$ which are associated with the spin connection part of $D_a$.
These operators are defined by $T_a := \frac{1}{4i}f_{abc}\gamma_{bc}$ and form a 16-dimensional representation of the $su(3)$ Lie algebra that we call the ``Clifford representation".

\be \label{C1}
[T_a , T_b] = if_{abc}T_c.
\ee
Observe that $\gamma_{a b c \cdots}$ transforms as a tensor in this representation. Objects with tensor field character like $P_{ab},J_{ab}, \gamma_{ab}$ transform as such under coordinate rotations with generators $T+ \mathcal{L}$ in the product representation\footnote{To prove the second equation here we used the fact that the $d$-tensor is $SU(3)$-invariant, this means that $$f_{abe}d_{ecd}+f_{ace}d_{bed}+f_{ade}d_{bce}= 0 $$ }: 

\beqa \label{C2}
\lbrack T_a +\mathcal{L}_a, \gamma_b \rbrack ~ = ~ \lbrack T_a, \gamma_b \rbrack &=& i f_{abc} \gamma_c , \nn \\
\lbrack T_a + \mathcal{L}_a, P_{bc} \rbrack ~ = ~ \mathcal{L}_a P_{bc}  & = & if_{abd}P_{dc} + if_{acd} P_{bd}.
\eeqa
It may be shown that the Clifford representation, $Cliff$, has a quadratic Casimir equal to that of the adjoint:

\be
C_2(SU(3),Cliff)= T_a T_a = \frac{1}{8} f_{abc}f_{abc} = \frac{1}{8}\Tr (C_2 ( SU(3),Adj)) = 3.
\ee
The quadratic Casimir in our convention is defined as the sum of the squares of the generators that satisfy (\ref{C1}) with the structure constants defined by the standard Gell-Mann matrices as in (\ref{A00}).
It is easy to convince oneself that this is enough information to determine the breaking of $Cliff$ into irreducible $SU(3)$ representations as the sum of two adjoints:

\be \label{C3}
Cliff = \young(\ \ ,\ )\oplus  \young(\ \ ,\ )  =  {\bf 8} \oplus {\bf 8} 
\ee

To calculate the spectrum of our Dirac operator we first compute its square, this should be sufficient since we know that for $\CP^N$ the Dirac operator has a symmetrical spectrum around zero\cite{Doc, spinor}:

\be \label{C4}
\D^{~2} = D_a D_a +\half \gamma_{ab} [D_a, D_b] -i \gamma_{ab}P_{ac}f_{cbe}D_e  
\ee 
to simplify the last expression we make first a few remarks on the orbit construction of $\CP^2$.
We are conceiving $\CP^2$ as a coset space $G/H$ with $G= SU(3)$ and $H=S(U(2) \times U(1))$; at each point there is an isotropy subgroup isomorphic to $H$ whose generators are linear combinations of the generators of $G$ with point dependent coefficients. Hence, we are led to a natural separation of the Lie algebra of $G$, denoted by $\underline{G}$, at each point, into the isotropy subalgebra $\underline{H}$ and its orthogonal complement $\underline{G/H}$:

\be \nn
\underline{G}= \underline{H} \oplus_{\perp} \underline{G/H}
\ee 
Making use of this fact we shall relabel the structure constants of the group at each point so that the indices

\beqa
a,b,c,d,e,\cdots \quad & \mbox{label} & \quad \underline{G} \nn \\
\alpha, \beta,\gamma ,\cdots \quad &\mbox{label}& \quad \underline{H} \nn  \\
i,j,k, \cdots \quad & \mbox{label}& \quad \underline{G/H} \nn
\eeqa
In our case we know that $G$ is compact and $H \subset G$ is a subgroup, therefore

$$
f_{\alpha \beta k} = f_{\alpha k \beta} = f_{k \alpha \beta}=0.
$$
It is a known fact that $\CP^2$ is a symmetric space\footnote{this brings as a consequence the vanishing of the torsion tensor $f_{ijk}$ for the canonical covariant derivative\cite{Bal2}} and this fact implies $f_{ijk}=0$, summarising:

\beqa  \label{C5}
\lbrack \underline{H}, \underline{H}\rbrack &  \subset & \underline{H} \nn \\
\lbrack \underline{H}, \underline{G/H}\rbrack &  \subset & \underline{G/H} \nn \\
\lbrack \underline{G/H}, \underline{G/H}\rbrack &  \subset & \underline{H} \nn
\eeqa
With these considerations we may rewrite (\ref{C4}) as

\be \nn
\D^{~2} = \Delta_s + \half \gamma \cdot F 
\ee
where the curvature of the spin connection has been introduced:

\be \nn
F_{ab} := [D_a , D_b] - if_{abc}D_c 
\ee
and the spin laplacian has been identified as the square of the covariant derivative $\Delta_s= D_a D_a$. The curvature term can be further simplified, by writing out the curvature tensor components explicitely 

\be \nn
F_{ab} = i [P_{ad}P_{be}f_{dec} + f_{abe}P_{ec}+ f_{ace}P_{be}-f_{bce}P_{ae}]T_c
\ee
one sees at once that due to (\ref{C5}) it has only tangent nonvanishing components $F_{jk}=-if_{jk \alpha}T_{\alpha}$. After some manipulations, together with the derivation of the scalar Ricci curvature $R$ given in appendix \ref{A}, we may show that the curvature term is just

\be \label{C6}
\half \gamma \cdot F = -\frac{1}{8} f_{ij \alpha}f_{\alpha kl}\gamma_{i}\gamma_{j}\gamma_{k}\gamma_{l}= \frac{R}{4} {\bf 1}
\ee 
In our convention for $\CP^2$ (\ref{A2}) we find the value $R=6$. Equation (\ref{C6}) shows that our ansatz is consistent with Lichnerowicz's theorem \cite{Lawson}. 

\section{Spin$_c$ bundle construction and spectrum } \label{spinspec}

In this section the spin$_c$ bundle is constructed and the spectrum along with all the eigenfunctions of the Dirac operator are found using representation theory. As it turns out, one obtains a perfect match with the spectrum of the known Dirac operator corresponding to canonical spin$_c$ structure on $\CP^2$, this is a hint that our ansatz for the spin connection corresponds to such choice. We prove that this is indeed the case by looking at the representation content of the spin bundle. Also, two equivalent forms of the chirality operator are given.

The first observation \cite{Brianx, Briany} is that we may rewrite the spin Laplacian in terms of quadratic Casimir operators of the groups involved, explicitely

\be \label{D0}
\Delta_s =(\mathcal{L}+T)^2 - T \cdot ( {\bf 1}- P) T = C_{2}(SU(3), \cdot) - C_{2}(S(U(2)\times U(1)), \cdot ) 
\ee
We still need to determine the relevant representations for which these Casimir operators are to be evaluated. This is most easily done if we first analyze which representations can occur as a result of the sum of ``angular momenta" $\mathcal{L}+T$.
A spinor field in this construction will be a function-valued 16-component column object, this is, $\psi \in S:= L^2 (\CP^2,\mathbb{C}) \otimes Cliff $. Since the Clifford representation is just a sum of two copies of the ${\bf 8}$ we may restrict our attention to the spin components $\Psi \in L^2 (\CP^2,\mathbb{C}) \otimes {\bf 8}$ and use the harmonic decomposition (\ref{B3}). Hence, the representations that will result from the sum $\mathcal{L}+T$ will be just the reduction of

\be  \label{D1}
 S= {\bf 8}\otimes ({\bf 1 }\oplus {\bf 8}\oplus {\bf 27 }\oplus{\bf 64 }\oplus{\bf 125 }\oplus \cdots )=(1,1) \otimes \bigoplus_{l=0}^{\infty}(l,l)
\ee
The $SU(3)$ content of a generic term in the series is then given by the Clebsch-Gordan decomposition:

\beqa  \label{D2}
(1,1) \otimes (l,l) = \underbrace{(l+1,l+1)}_{l \geq 0}\oplus \underbrace{(l+2,l-1) \oplus (l-1,l+2) \oplus (l,l) \oplus (l,l)}_{l \geq 1} \\ \nn
\oplus \underbrace{(l+1,l-2) \oplus (l-2,l+1)}_{l \geq 2}\oplus \underbrace{(l-1,l-1)}_{l \geq 1}
\eeqa
The restrictions placed below tell when this particular representations appear in the series.
We want to project onto specific components of the Clebsch-Gordan series in order to obtain definite values of the $SU(3)$ Casimir operator to construct eigenspinors. For this we use the $SU(3)$ Clebsch-Gordan coefficients, (see \cite{Swart} for details, the $SU(3)$ can be constructed from those of $SU(2)$ and isoscalar factors)

\be
\langle (l_1^{\prime} , l_2^{\prime}), I' ,I_3^{\prime}, Y' ; (l_1^{\prime \prime} , l_2^{\prime \prime}), I'' ,I_3^{\prime \prime}, Y'' |  (l_1 , l_2)_{R}, I ,I_3, Y \rangle := \left(\begin{array}{ccc} (l_1^{\prime},l_2^{\prime}) &(l_1^{\prime \prime},l_2^{\prime \prime}) & (l_1, l_2)_{R} \\
I' I_3^{\prime} Y' & I'' I_3^{\prime \prime} Y'' & I I_3 Y
\end{array}\right) \nn
\ee
Where the pair $(l_1 ,l_2)$ are the Dynkin indices and the extra subindex $R$ is added to distinguish identical representations in the Clebsch-Gordan series. In what follows $e_a$ represents a basis vector of an ${\bf 8}$ in the decomposition (\ref{C3}). Only certain representations interest us, those associated with the spinors on $\CP^2$. The field $\Psi$ must be projected onto the appropriate subspace. We later show that the representations which give rise to this subspace are (together with an harmonic spinor $\Phi_0$)

\beqa \label{D3}
\Phi^{(l-2,~l+1)}_{II_3 Y} & := & \sum_{\begin{subarray}{1} 
I',I_3^{\prime},Y' \\ 
~~a 
\end{subarray}} \left( \begin{array}{ccc}
(l,l) & (1,1) & (l-2,l+1) \\
I' I_3^{\prime} Y' & a & I I_3 Y
\end{array}\right) \mathcal{Y}_{I' I_3^{\prime} Y'}^{l} \otimes e_a \quad \quad l \geq 2  \nn \\      \!\!\!\Phi^{(l,l)}_{II_3 Y} & := & \sum_{\begin{subarray}{1} 
I',I_3^{\prime},Y' \\ 
~~a 
\end{subarray}} \left( \begin{array}{ccc}
(l,l) & (1,1) & (l,l)_{R} \\
I' I_3^{\prime} Y' & a & I I_3 Y
\end{array}\right) \mathcal{Y}_{I' I_3^{\prime} Y'}^{l} \otimes e_a \quad \quad \quad \quad \quad l \geq 1   \\ \nn 
\eeqa
Now we further decompose ${\bf 8}$ into irreducible representations of the subgroup $S(U(2)\times U(1))$ through the branching rule for the fundamental representation (writing the $U(1)$ charge as a subindex):

\be \nn
\young(\ ) = {\bf 3 } = {\bf 2}_{1} \oplus {\bf 1}_{-2}
\ee
whence
\be \nn
\young(\ ,\ ) = \bar{ {\bf 3}} = {\bf 2}_{-1} \oplus {\bf 1}_{2}
\ee
is obtained by conjugation. Tensoring up these representations yields

\be  \label{D4}
\young(\ \ ,\ )= {\bf 8} ={\bf 2}_{3} \oplus {\bf 3}_{0} \oplus {\bf 1 }_{0}\oplus {\bf 2}_{-3}
\ee
To distinguish the relevant representations, one needs an operator that distinguishes the $U(1)$ charges of the fields involved. The generator of the $U(1)$ isotropy subgroup may serve for this purpose; its eigenvalues distinguish the charges:

\be \nn
\phi := \frac{1}{4i} J_{ab} \gamma_{ab} = \frac{2}{\sqrt{3}}\xi \cdot T.
\ee
Evaluating it at the ``north pole" gives

\be \nn
\phi^0 = - \frac{2}{\sqrt{3}} T_8.
\ee
In the appendix \ref{B} it is shown that it has the minimal polynomial

\be \nn
\phi (\phi^2 - 1) = 0 
\ee
and hence eigenvalues $0, \pm 1$ associated with the charges of (\ref{D4}). With this information we may construct the projectors from ${\bf 8}$ onto ${\bf 2}_{\pm 3}$ in a canonical manner: $\half \phi (\phi \pm 1)$
There is also a canonical chirality operator in the background space $\mathbb{R}^8$, $\gamma$ given by:

\be \nn
\gamma := \prod_{a=1}^8 \gamma_a, \quad \quad \gamma^2 = 1, \quad \{ \gamma_a ,\gamma \}=0
\ee
After $\{ \gamma , \D ~\}=0$ we would like to identify $\gamma$ as the chirality operator for $\D$, it will be proven that when properly restricted, $\gamma$ coincides with the chirality.
It may also be proven from the definition, the anticommutation relations and (\ref{C2}) that

\be \nn
 \quad \quad [T_a , \gamma ] = 0  ~~\left( \mbox{indeed}~~[\mathcal{L}+T, \gamma]=0 \right)
\ee
This equation shows that the breaking (\ref{C3}) is respected by the chirality, therefore each tensor in ${\bf 8}$ will have a definite chirality $\pm 1$ according to $\gamma$ and we use this property to construct the projectors onto the two copies ${\bf 8}$ that conform the Clifford representation. The fact that $\gamma$ is an $SU(3)$ scalar implies that the components belonging to different ${\bf 8}$'s do not mix under $SU(3)$ rotations. If $f$ denotes the basis vectors of the Clifford representation we obtain the two sets of $e_a$, one for each ${\bf 8}$, say $e^{\pm}$, given as $e^{\pm} =\frac{1}{2}(1 \pm \gamma) f$.
This information is now enough to build the projectors onto the representations ${\bf 2}_{\pm 3}$
(notice $[\phi,\gamma]=0$) :

\be \nn
\pi_{\pm}= \frac{1}{4} \phi (\phi \pm 1)(1 \pm \gamma) ~~~\mbox{and}~~~\tilde{\pi}_{\pm}= \frac{1}{4} \phi (\phi \mp 1)(1 \pm \gamma)
\ee
There are four projectors $\pi_{\pm}$ and $\tilde{\pi}_{\pm}$ since there are two copies of the ${\bf 8}$. For future convenience we will focus on the copy with negative chirality. Observe that $\Tr \pi_{\pm} = \Tr \tilde{\pi}_{\pm} = 2$ follows from (\ref{x11}). We remark that $\pi_{\pm}, \tilde{\pi}_{\pm},\phi, \gamma$ are all $SU(3)$ scalars.

\subsection{Construction of the spin$_c$ bundle}

Our proposed spin$_c$ bundle is defined to be 

\be \label{D5}
\mathfrak{S}  = im (\pi_-) \oplus im(\D ~ \pi_-) \oplus K
\ee
From $\pi_- \D ~ \pi_- = 0$ one sees at once that $im (\pi_-) \cap im(\D ~ \pi_-) = 0$ and a little thought shows that $ker (\D ~)\cap im(\pi_-)=0$, because otherwise $\D^{~2}$ would have an harmonic spinor (zero mode) in $im(\pi_-)$( see lemma 1 below). The space $K \subset ker(\D~)$ is generated by the one harmonic spinor $\Phi_0$ present in the canonical spin$_c$ structure (see lemma 2 for definition), hence $K \cap im(\pi_-)= 0$ and $K \cap im(\D~ \pi_-)$ by lemma 1, hence $\mathfrak{S}$ has dimension 4 over $L^2 ( \CP^2,\mathbb{C})$.
It is also clear from the definition and the $SU(3)$-invariance property of $\pi_-$ that $\mathfrak{S}$ is invariant under $\D$ and $im(\pi_-)$ is invariant under $\D~^2$ but not under $\D~$.

Notice from (\ref{C0}) that a field which is an $SU(3)$ scalar lies in $ker(\D~)$.

\flushleft
\newtheorem{qqa}{Lemma}
\begin{qqa}
$ker(\D~^2) \cap im(\pi_-) = 0$ 
\end{qqa}

\vspace{0.2cm}
{\it Proof:} It suffices to prove that the spin laplacian $\Delta_s$ is positive definite in $im(\pi_-)$. Using (\ref{D6}) and the fact that the minimum value of the $SU(3)$ quadratic Casimir in $im(\pi_-)$ is, from (\ref{w8}), just 3, the value of the spin laplacian is at least $\frac{3}{2}$ on $im(\pi_-)$ $\blacksquare$

\flushleft
\newtheorem{qqa1}[qqa]{Lemma}
\begin{qqa1}
All harmonic fields found in $S$ of the type $(0,0)$ are multiples of $\xi \cdot e^{\pm}$. 
\end{qqa1}

\vspace{0.2cm}
{\it Proof:} The representation $(0,0)$ only appears once in (\ref{D2}), as a result of the product ${\bf 8} \otimes {\bf 8}$ in each copy, the base vectors for such representations are $\xi_a$ and $e_a^{\pm}$, whose only $SU(3)$ scalars are multiples of $\xi \cdot e^{\pm}$. We will define $\Phi_0 := \xi \cdot e^{+}$ and justify this choice at the end of the next subsection  $\blacksquare$

\vspace{0.2cm}

The chirality operator is defined as the product of all tangent gamma matrices at each point:

\be \nn
\Gamma  := \prod_{i} \gamma_{i} = \frac{1}{8} J_{ab}J_{cd}\gamma_{abcd}
\ee
With this choice of the phase one has at the ``north pole" $\Gamma^0 = \gamma_{4}\gamma_{5}\gamma_{6}\gamma_{7}$. Also from (\ref{x10}),

\be
\Gamma = 1 - 2 \phi^2 , ~~~~~~ \Gamma \pi_-  = -\pi_-
\ee
showing that $im(\pi_-)$ has negative chirality according to $\Gamma$. It is easy to check that $\{ \D~ , \Gamma \} =0$ (appendix \ref{C}), then $im(\D~\pi_-)$ has positive chirality according to $\Gamma$. The chirality defined is an $SU(3)$ scalar, it satisfies $\Gamma^2 ={\bf 1}$ and leaves $\mathfrak{S}$ invariant.

\subsection{Identifying the canonical spin$_c$ bundle}

We collect some results aimed to prove that the representation content of our proposed spin$_c$ bundle $\mathfrak{S}$, is the same as that of the sections for the canonical spin$_c$ bundle from the literature, and to connect our construction with the standard formulation. 

\vspace{-0.2cm}

\flushleft
\newtheorem{zza}{Assertion}
\begin{zza}
The space $im(\D ~ \pi_-)$ has the same $SU(3)$ representation content as $im(\pi_-)$.
\end{zza}

\vspace{0.2cm}

{\it Proof:} The Dirac operator $\D~$ is an $SU(3)$ scalar, for it commutes with $\mathcal{L}+T$, hence $\D~$ does not affect the transformation properties of the fields. Clearly $\D~$ does not annihilate any representation because this would lead to $ ker(\D^{~2}) \cap im(\pi_-) \neq 0$, in contradiction to lemma 1, consequently no representations will be missing  $\blacksquare$

\newtheorem{zzb}[zza]{Assertion}
\begin{zzb}
$\phi \D~ \pi_- =0$.
\end{zzb}

\vspace{0.2cm}

{\it Proof:} We prove the stronger result $\phi \D~ \phi = 0$. First  $\phi \D~ \phi =  \phi \gamma_a P_{ab} (\mathcal{L} + T)_b \phi = \phi \gamma_a P_{ab} \phi (\mathcal{L} + T)_b $.  Using the definitions in appendix \ref{C}: $ \phi= \mathfrak{g} \phi^0 \mathfrak{g}^{\dagger}$ and being $\xi_b = D_{ab} \xi_a^0$ we find $\phi \gamma_a P_{ab} \phi = \mathfrak{g} \phi^0 \gamma_a P^0_{ac} \phi^0 \mathfrak{g}^{\dagger} D_{cb}$ and it is easily verified that $\phi^0 \gamma_a P^0_{ac} \phi^0 =0$  $\blacksquare$

\newtheorem{wwa}{Corollary}
\begin{wwa}
$(1-\Gamma)\D~ \pi_- = 0$.
\end{wwa}

\vspace{0.2cm}
{\it Proof: } This is trivial from $(1-\Gamma)= 2 \phi^2$ and assertion 2.  $\blacksquare$

\newtheorem{zzd}[zza]{Assertion}
\begin{zzd}
$im (\D ~ \pi_-) \subset {\bf 1}_0 \oplus{\bf 3}_0$.
\end{zzd}

\vspace{0.2cm}
{\it Proof:} Assertion 2 shows that fields on $im(\D ~ \pi_-)$ have zero $U(1)$ charge. They also have a definite positive chirality since $\gamma \D ~ \pi_- = - \D ~ \gamma \pi_- =  \D ~ \pi_-$  $\blacksquare$

 Notice that $im(\D~ \pi_-)$ has also a positive chirality according to $\Gamma$, we will see below that $\gamma$ and $\Gamma$ agree on $\mathfrak{S}$. From (\ref{x3}) in appendix \ref{B} we may prove that the projectors from $\mathfrak{S}$ onto the ${\bf 1}_0 , {\bf 3}_0$ subspaces are indeed given by products of differences of quadratic Casimir operators:

\be \nn
P_{1_0} = (1-\phi^2)(1-\frac{C_2 (S(U(2)\times U(1)))}{2})  \quad  P_{3_0} =(1-\phi^2) \frac{C_2 (S(U(2)\times U(1)))}{2}
\ee
The spinor fields on $\CP^2$ admit the following decomposition \cite{Witten} (valid for arbitrary spin$_c$ structures):

\be \nn
| \psi \rangle  = \psi_0 | \Omega \rangle +  \psi_{\bar{\j}} \gamma^{\bar{\j}} | \Omega \rangle +  \psi_{\bar{1} \bar{2}} \gamma^{\bar{1} \bar{2}} | \Omega \rangle,  \quad \quad \bar{\j} = \bar{1} ,\bar{2} 
\ee
Where $ | \Omega \rangle $ is the vacuum annihilated by the holomorphic set of gamma matrices $\gamma^{\j} = (\gamma^{\bar{\j}})^{\dagger}$. The gamma matrices in this setting are distinguished as holomorphic and antiholomorphic, their defining relations being:

\be \nn
\{ \gamma^{\i} ,\gamma^{\bar{\j}} \} = \delta^{\i \bar{\j}} {\bf}, \quad \{ \gamma^{\i}, \gamma^{\j} \} =  \{ \gamma^{\bar{\i}}, \gamma^{\bar{\j}} \} = 0
\ee

Since $\CP^2$ is not Calabi-Yau the component $\psi_0$ has a $U(1)$ charge which is compensated exactly by the charge from the spin connection in the canonical spin$_c$ structure, and thus the charge of each contribution for $|\psi \rangle$  vanishes although the charges of the individual components differ (This means that the spinors may be identified with ordinary $(0,k)$-forms on $\CP^2$ \cite{spinor, Witten} ).

\nin The $SU(3)$ content for each component of the canonical spin$_c$ bundle is known to be \cite{spinor},

\beqa \nn
| \psi \rangle \in
\Bigg(\underbrace{\bigoplus_{l=1}^{\infty}(l,l)\bigoplus_{l=2}^{\infty}(l-2,l+1)}_{ \psi_{\bar{\j}}} \Bigg)&\underbrace{\bigoplus_{l=0}^{\infty}(l,l)}_{\psi_0}& \Bigg( \underbrace{\bigoplus_{l=2}^{\infty} (l-2,l+1)}_{\psi_{\bar{1} \bar{2}}} \Bigg) \\ \nn
\eeqa
The harmonic spinor that generates $K$ belongs to the representation $(0,0)$ (behaves as a scalar), has positive chirality and is denoted by $\Phi_0$.
We show in section \ref{F6Y} that the representation content of $im (\pi_-)$ is

\be \label{D6}
im (\pi_-) = \bigoplus_{l=1}^{\infty} (l,l) \bigoplus_{l=2}^{\infty} (l-2,l+1)
\ee
We can now see that the representation content of our proposed spin$_c$ bundle $\mathfrak{S}$ is identical to the known for the canonical spin$_c$ bundle, identifying $\mathfrak{S}$ as the forementioned bundle. 

By choosing the harmonic spinor to have positive chirality, $\Phi_0  = \xi \cdot e^+  \in {\bf 1_0}$, we can conclude that  actually $\psi_{\bar{\j}} \in im (\pi_-)$. The index of our Dirac operator, restricted to $\mathfrak{S}$, is in agreement with known results since $ind (\D~|_{\mathfrak{S}}) = 1$.
The structure of our spin$_c$ bundle can be summarized as follows: one piece with no zero modes and negative chirality $im(\pi_-) = {\bf 2}_{-3}$, multiples of one harmonic spinor (zero mode) with positive chirality, $\alpha \Phi_0 \in K \subset {\bf 1}_0$, and one piece with no zero modes and positive chirality, $im(\D~ \pi_-) \subset {\bf 1}_0 \oplus {\bf 3}_0$. To end this section we prove the following
\flushleft
\newtheorem{qqb}[zza]{Assertion}
\begin{qqb}
 If we restrict the chirality to act upon $\mathfrak{S}$ then $\Gamma|_{\mathfrak{S}} = \gamma|_{\mathfrak{S}}$.
\end{qqb}

\vspace{.2cm}

{\it Proof:} Let $\Psi = \pi_- \psi_1 + \D~\pi_- \psi_2 + \alpha \Phi_0 \in \mathfrak{S}; ~ \psi_{1},\psi_2 \in S, ~ \alpha \in \mathbb{C}$ be a general element, then 

\beqa \nn
\gamma \Psi & = & \gamma \pi_- \psi_1 + \gamma \D~\pi_- \psi_2 + \alpha \gamma \Phi_0 \\ \nn
            & = & - \pi_- \psi_1 +  \D~ \pi_- \psi_2 + \alpha \Phi_0 
\eeqa 
but also

\beqa \nn
 \Gamma \Psi & = & \Gamma \pi_- \psi_1 + \Gamma \D~\pi_- \psi_2 + \alpha \Gamma \Phi_0 \\ \nn
             & = & -\pi_- \psi_1 - \D ~ \Gamma \pi_- \psi_2 + \alpha \Gamma \Phi_0 \\ \nn
            & = & - \pi_- \psi_1 +  \D~ \pi_- \psi_2 + \alpha \Gamma \Phi_0 
\eeqa
It can be seen that $\Gamma \Phi_0 = \Phi_0$, since $\phi^2 \Phi_0 = 0$. Indeed $\phi \Phi_0 = 0$ follows from $ \phi \sim \xi \cdot T$ and, in a certain base $T_a e_b^+ = -if_{abc} e_c^+$ implies that $\phi \Phi_0 \sim \xi_a \xi_b f_{abc} e_c^+ = 0$  $\blacksquare$

\subsection{Spectrum of the Ansatz Dirac operator}

We will later show that $ im( \pi_-)$ is spanned by the functions (\ref{D3}). We leave for appendix \ref{C} the proof that $C_{2}(S(U(2) \times U(1)))$ has a definite value on $im (\pi_-)$,

\be \label{D7}
C_{2}(S(U(2) \times U(1))) \pi_- = \frac{3}{2} \pi_-
\ee

A basis of eigenspinors for $\D$ is then given by adequate projections of the functions (\ref{D3}), 

\beqa \label{D8}
\Psi^{(l-2,~l+1)}_{II_3 Y ; \pm} &=& \big( \pi_{-} \pm \frac{1}{\sqrt{l(l+1)}}\D~ \pi_- \big)   \Phi^{(l-2,~l+1)}_{II_3 Y} \quad l \geq 2 \nn \\
\Psi^{(l,~l)}_{II_3 Y ; \pm}&=&  \big( \pi_{-} \pm \frac{1}{\sqrt{l(l+2)}}\D~ \pi_- \big) \Phi^{(l,~l)}_{II_3 Y} \quad l \geq 1,  \\ \nn
\eeqa
together with the harmonic spinor $\Phi_0$. The invariance of $\pi_{-}$ and formulae (\ref{C6}), (\ref{D0}), (\ref{D7}) allows one to calculate the spectrum of $\D^{~2} ~|_{im(\pi_-)}$, which gave the eigenspinors (\ref{D8}),

\beqa \nn
\D^{~2}\pi_- \Phi^{(l-2,~l+1)}_{II_3 Y }&=& l(l+1) \pi_- \Phi^{(l-2,~l+1)}_{II_3 Y } \\ \nn
\D^{~2} \pi_- \Phi^{(l,~l)}_{II_3 Y} &=& l(l+2) \pi_- \Phi^{(l,~l)}_{II_3 Y}  
\eeqa
The spectrum of $\D~^2|_{im(\pi_-)}$ is given below, it can be calculated following appendices \ref{B}, \ref{C} and the remarks just made:

\be \nn
\mbox{Spec}\{\D^{~2}|_{im (\pi_-)} \}= \{l(l+2) : l \in \mathbb{N}  \} \cup \{ l(l+1) : l \in \mathbb{N}-\{ 1 \}  \}
\ee
each with degeneracy 

\beqa 
\mbox{deg}~l(l+2) &=& (l+1)^3 ~ = ~ \mbox{dim}(l,l) \nn \\
\mbox{deg}~l(l+1) &=& \frac{(2l+1)(l-1)(l+2)}{2} ~=~ \mbox{dim}(l-2,l+1) \nn 
\eeqa
The spectrum matches with results obtained in \cite{spinor, Grosse}. Finally, the spectrum of $\D ~|_{\mathfrak{S}}$ is symmetrical around zero, with the degeneracies mentioned (and deg(0) = 1):

\be \nn
\mbox{Spec}\{\D ~|_{\mathfrak{S}} \}= \{ \pm \sqrt{l(l+2)} : l \in \mathbb{N}  \} \cup \{ \pm \sqrt{l(l+1)} : l \in \mathbb{N}/\{ 1 \}  \}\cup \{ 0 \}
\ee
the corresponding eigenspinors are

\beqa \nn
\D ~\Psi^{(l+2,~l-1)}_{II_3Y; \pm} &=& \pm (l+1)(l+2) \Psi^{(l+2,~l-1)}_{II_3Y; \pm} \\ \nn 
\D ~\Psi^{(l,~l)}_{II_3Y; \pm} &=& \pm l(l+2) \Psi^{(l,~l)}_{II_3Y; \pm} \\ \nn
\D~ \Phi_0 &=& 0
\eeqa

\section{Fuzzy Construction} \label{F6Y}

To achieve a fuzzy version of the spinors it is necessary to substitute the algebra of functions by a sequence of finite dimensional algebras that in the commutative limit recovers the usual algebra of functions.
The algebra of functions is to be replaced by the sequence of square matrix algebras $(0,L)\otimes (L,0)$, $S$ then becomes $(0,L) \otimes (L,0) \otimes (1,1)$. Consider its decomposition into irreducible representations by steps, first taking the product $(L,0)\otimes (1,1)$

\be \label{E0}
(L,0)\otimes (1,1) = (L+1,1)\oplus (L-1,2) \oplus (L,0) \oplus (L-2,1):=S_1 \oplus S_2 \oplus S_3 \oplus S_4.
\ee
This produces four projective (left) modules over the ring $(0,L) \otimes (L,0)$ that we identify as the fuzzy version of (\ref{D4}). Further reduction of each module represents the harmonic decomposition of the module, very much like (\ref{B3}) is for a function.
The projective modules have, in the commutative large $L$ limit, the corresponding dimensions of the representations (\ref{D4}). To count the dimensions of the subbundles that we are obtaining in their fuzzy version we must divide out the dimensions of the right hand side in (\ref{E0}) by the dimension of $(L,0)$, to factor out the functional degrees of freedom, and take the large $L$ limit.

\beqa \nn
\mbox{dim}S_{1} &=&  (L+2)(L+4) \\ \nn
\mbox{dim}S_{2} &=&  \frac{3}{2}L(L+3)  \\ \nn
\mbox{dim}S_{3} &=&  \frac{1}{2}(L+2)(L+1)  \\ \nn
\mbox{dim}S_{4} &=&  L^2-1 \\ \nn
\eeqa
giving for the limiting quotient ratios:

\beqa \nn
\lim_{L \to \infty} \frac{\mbox{dim}S_{1}}{\mbox{dim}(L,0)} = 2&~&~~ \lim_{L \to \infty} \frac{\mbox{dim}S_{2}}{\mbox{dim}(L,0)} = 3\\ \nn
\lim_{L \to \infty} \frac{\mbox{dim}S_{3}}{\mbox{dim}(L,0)} =1 &~&~~ \lim_{L \to \infty} \frac{\mbox{dim}S_{4}}{\mbox{dim}(L,0)} =2. \\ \nn
\eeqa
in agreement with our interpretation.

The important remark is that if one analyses the harmonic decomposition of these representations it is precisely $S_4$ that gives in the large $L$  limit the series of $SU(3)$ needed for the spinors discussed before, that is, it gives the exact $SU(3)$ representation content of (\ref{D3}). For $L \geq 2$ the harmonic decomposition of the fuzzy ``subbundles" (our projective modules) reads:

\beqa \nn
(0,L)\otimes S_1 &=&\bigoplus_{l=1}^{L+1} (l,l)\bigoplus_{l=1}^{L} (l+2,l-1)\\ \nn
(0,L)\otimes S_2 &=&\bigoplus_{l=1}^{L} (l-1,l+2)\bigoplus_{l=1}^{L} (l,l)\bigoplus_{l=2}^{L} (l+1,l-2)\\ \nn
(0,L)\otimes S_3 &=&\bigoplus_{l=0}^{L}(l,l)\\ \nn
(0,L)\otimes S_4 &=&\bigoplus_{l=2}^{L} (l-2,l+1)\bigoplus_{l=1}^{L-1} (l,l) 
\eeqa
We wish to construct the projections over the projective modules $S_1$ and $S_4$, this is done by the standard technique using quadratic Casimir operators \cite{Julieta, Master}

\beqa \nn
\hat{P}_1 &=& \frac{(L \hat{\phi} +1)(L \hat{\phi}+3)(L \hat{\phi}+L+3)}{(L+1)(L+3)(2L+3)}\bigg( \frac{1-\gamma}{2} \bigg) \\ \nn 
\hat{P}_4 &=& \frac{(1-\hat{\phi})(L \hat{\phi}+1)(L \hat{\phi}+3)}{(2L+3)(L+2)}\bigg( \frac{1-\gamma}{2} \bigg)
\eeqa
where the fuzzy analogue of $\phi$, called $\hat{\phi}$, has been introduced as the operator

\be \nn
\hat{\phi} = \frac{2}{L}{\bf L}\cdot T.
\ee
This operator has $\phi$ as its commutative limit and has the minimum polynomial

\be \nn
(\hat{\phi}-1)(\hat{\phi}+1 +\frac{3}{L})(\hat{\phi}+\frac{1}{L})(\hat{\phi}+\frac{3}{L})=0.
\ee
Using these relations it is very easy to find the commutative limit of the projectors $\hat{P}_1$ and $\hat{P}_4$, they are

\be \nn
P_1 =  \frac{1}{4} \phi (\phi+1)(1-\gamma) = \tilde{\pi}_-, \quad \quad P_4 =  \frac{1}{4} \phi (\phi-1)(1-\gamma)=\pi_-
\ee
Indeed $im(\pi_-) = im(P_4)$ is spanned by the functions (\ref{D3}) projected onto one ${\bf 8}$, as claimed.

\section{Conclusions}
In this paper we have proposed an ansatz for the projective Dirac operator on $\CP^2$, it turned out that our ansatz corresponds to the canonical choice of spin$_c$ structure. We have calculated the spectrum of our Dirac operator and constructed the spinors whereupon it acts, the obtained spectrum and eigenspinors are in agreement with our interpretation. A novel feature of this construction is that it does not make reference to any local coordinate system (as in classical differential geometry) but rather uses the global embedding coordinate system from \cite{CPNF} compatible with fuzzy complex projective spaces $\CP^N_F$. This construction brings nearer the goal of obtaining a fuzzy QED theory on a 4-dimensional space, namely $\CP^2_F$. As a result of our choice of the coordinate system we had to reduce the total spinor space, $S$, to an appropriate physical subspace $\mathfrak{S}$. A fuzzy analogue for $S$ was used to find the representation content of the relevant subbundles of $\mathfrak{S}$, which were then related to the standard construction of the spin$_c$ bundle. It would be interesting to generalise this work by including other spin$_c$ structures, higher dimensions, extension to Grassmann manifolds or to continue towards QED on $\CP^2_F$.

\vspace{1cm}

{\Large{ \bf Acknowledgments}} 

\vspace{0.5cm}

The author is grateful to IFM, UMSNH where part of this work was carried out and to CONACyT, Mexico, for continued financial support. 
It is a pleasure to thank D. O'Connor and A. P. Balachandran for suggesting this problem, B. P. Dolan and E. Wagner for enlightening discussions, and to the TPI, Jena, for hospitality and support. My deepest thanks to A. I. Garc\'ia L\'opez for spotting a combinatorial error in the branching rule given in appendix \ref{C} in the previous version, and which has now been corrected.

\newpage
\begin{appendix}
\section{Curvature on reductive coset spaces} \label{A}

This appendix contains the calculation of the Riemann curvature tensor on homogeneous coset reductive spaces and the formula for the quadratic Casimir operators of $SU(3)$ to evaluate the Ricci scalar in our case of interest. The material of this appendix is not new at all, we follow \cite{Brianx, Salam, Popov}.

The homogeneous coset space $G/H$ of a connected Lie group $G$ of order $|G|$ is called reductive if it is possible to break the Lie algebra $\underline{G}$ as in (\ref{C5}) this is always the case when $G$ is compact and $H \subset G$. On every coset space $G/H$ there is a canonically induced $G$-invariant metric for which the generators of $\underline{G}$ are Killing vectors.

A set of Vielbeins for this metric can be constructed from the canonical Maurer-Cartan 1-forms on $G$ as we shall see. The Maurer-Cartan 1-form is a Lie algebra-valued 1-form on $G$ given by

\be \nn
\theta = g^{-1}dg~~~~~g \in G, 
\ee
and satisfies the Maurer-Cartan equation

\be \label{w1}
d \theta + \theta \wedge \theta =0,~~~~~~\theta=\theta^a \mathfrak{j}_a \in \underline{G}
\ee
since

\beqa \nn
d\theta &=& d(g^{-1} dg)= dg^{-1}\wedge dg = -g^{-1}dg g^{-1}\wedge dg \\ \nn
&=& -g^{-1}dg \wedge g^{-1}dg = - \theta \wedge \theta \nn
\eeqa
wherein we have denoted by $\mathfrak{j}_a$ the generators of $\underline{G}$ and $\theta^a$ are 1-forms on $G$.

\nin Writing (\ref{w1}) componentwise we might appreciate better this equation

\be \label{w2}
d \theta^a + \frac{1}{2}f^{a}_{~bc}\theta^b \wedge \theta^c =0.
\ee
The Vielbeins are given as $ie^a=\theta^a$ and are associated with a non-coordinate dual basis of the cotangent space, strictly speaking only $e^k$ are Vielbeins and the rest $e^{\alpha}$ are, however, linear functions of $e^k$ whose exact dependance is irrelevant for our concerns.

\nin The torsion and curvature 2-forms determine the torsion and curvature of a manifold, they are given through the Cartan structural equations:

\beqa \label{w3} 
de^i+\omega^{i}_{~k}\wedge e^k&=&T^i:=\frac{1}{2}T^{i}_{~lm}e^l\wedge e^m \\   \label{w3.5} d\omega^{i}_{~k}+\omega^{i}_{~j}\wedge \omega^{i}_{~k}&=&R^{i}_{~k}:= \frac{1}{2} R^{i}_{~klm} e^l\wedge e^m
\eeqa
the quantities $T^{i}_{~lm}$ and $R^{i}_{klm}$ are the torsion and Riemann curvature tensors respectively.

\nin The Levi-Civita connection is unique on $G/H$, it is compatible with the induced $G$-invariant metric and has vanishing torsion. 

\nin Setting $T^i =0$ in (\ref{w3}) and comparing with (\ref{w2}) using the relationship between $e$ and $\theta$ one finds

\be \nn
de^i = \frac{1}{2} f^{i}_{~jk} e^j \wedge e^k + f^{i}_{~\alpha k} e^{\alpha}\wedge e^{k}
\ee
it is immediate that

\be \label{w4}
\omega^{i}_{~k} = \frac{1}{2}f^{i}_{~jk} e^j + f^{i}_{~k \alpha } e^{\alpha}.
\ee
We then calculate $R^i_{~k}$ using (\ref{w3.5}) and (\ref{w4}), further simplification thanks to the Jacobi identity and $f^{\alpha}_{i \beta}=0$ results in

\be
R^{i}_{~k} = \frac{1}{4}( 2 f^{i}_{~k \alpha}f^{\alpha}_{~lm}+f^{i}_{~kj}f^{j}_{~lm}-f^{i}_{~lj}f^{j}_{~km} )e^l \wedge e^m.
\ee
The Riemann curvature tensor is obtained from (\ref{w3.5})

\be \label{w5}
R^{i}_{~klm} = \frac{1}{2}( 2 f^{i}_{~k \alpha}f^{\alpha}_{~lm}+f^{i}_{~kj}f^{j}_{~lm}-f^{i}_{~lj}f^{j}_{~km} ).
\ee
A case of particular importance for us is when $G/H$ is a symmetric space, this means that the relation

\be \nn
[\underline{G/H},\underline{G/H}] \subset \underline{H}
\ee
holds, or equivalently $f^{i}_{~jk}=0$.
For our case of interest, $\CP^N$, this holds true and we are indeed dealing with a family of symmetric spaces. Formula (\ref{w5}) is then reduced to

\be \nn
R^i_{\mbox{\tiny{Symm}}~klm}=f^i_{~k \alpha }f^{\alpha}_{~lm}.
\ee 
The scalar Ricci curvature in this case is just

\be \label{w6}
R_{\mbox{\tiny{Symm}}} = f^{i}_{~k \alpha} f^{\alpha}_{~ik} = \frac{1}{3} \left( \Tr C_2 (G, Ad)- \Tr C_2(H, Ad)  \right).
\ee
For $\CP^N$, which is symmetric, with $G=SU(N+1)$ compact, one obtains through (\ref{w6}) 

\be \label{w7}
R_{\CP^N} = \frac{1}{3}[ ( (N+1)^2 -1 ) (N+1)-N(N^2-1) ]=N(N+1).
\ee
If we want to include the scale of the space so that the curvature tensor has the physical dimension of inverse area one should divide by the square of the ``radius"

\be \nn
R_{\CP^N} \longrightarrow \frac{N(N+1)}{\xi_a \xi_a}.
\ee
In calculating (\ref{w7}) we used the result

\be \nn
C_2 (S(U(N)\times U(1)),Ad) = N~ \boldsymbol{1}_{N^2-1} = C_2 (SU(N),Ad)
\ee
obtained from the Fierz identities for $SU(N)$ and analyzing the structure constants.

\nin Some useful formulae for the quadratic Casimir operators of $SU(N)$ are

\beqa \label{w8} 
C_2^{SU(3)}(l_1,l_2) &=& \Big[ \frac{1}{3}(l_1^2 + l_2^2 + l_1 l_2)+l_1+l_2 \Big]~ \boldsymbol{1} \\ \nn
C_2^{SU(N)}(Ad)&=&N ~\boldsymbol{1} \\ \nn
C_2^{SU(N)}(\mbox{\tiny{$\young(\ )$}})&=& \frac{N^2-1}{2N}~\boldsymbol{1}. \nn
\eeqa
The convention to define the Casimir operators is the one we fixed before.

\nin Formulae (\ref{w8}) are particular cases of the general quadratic Casimir of $SU(N)$ which can be found in \cite{Popov}.
Another useful fact is that the dimension of the irreducible representations of $SU(3)$ is given by

\be \nn
\mbox{dim} (l_1,l_2)= \frac{(l_1+l_2+2)(l_1 +1)(l_2+1)}{2}
\ee

\section{Minimum polynomial of the hypercharge}  \label{B}

The purpose of this appendix is to present a method to compute the minimum polynomial (and hence eigenvalues) of the $SU(3)$ hypercharge operator on a given point of $\CP^2$ defined as

\be \label{x1}
\phi = \frac{1}{4i} J_{ab} \gamma_{ab}
\ee
in terms of the complex structure $J_{ab}$ on $\CP^2$. We solve the problem for $\CP^N$ and then set $N=2$. The operator $\phi$ is proportional to the generator of the $U(1)$ in the isotropy subgroup $S(U(N)\times U(1))$ at any given point of the $\CP^N$, for instance in the ``north pole" defined above  we have the last generator of $SU(N+1)$ or hypercharge in the Clifford representation:

\be \label{x2}
\phi_{\mbox{\footnotesize{north pole} }} = - \sqrt{\frac{2N}{N+1}} T_{N^2 +2N}.
\ee
Hence, $\phi$ allows us to write the quadratic Casimir of $S(U(N)\times U(1))$ in a representation $\boldsymbol{m}_Q$ with $U(1)$ charge $Q$, that arises from the breaking of the Clifford representation, as

\be \label{x3}
C_2^{S(U(N)\times U(1))} (\boldsymbol{m}_Q) = C_{2}^{SU(N)}(\boldsymbol{m}) + \frac{N+1}{2N} \phi^2 |_{Q}
\ee
This Casimir operator is needed to compute the spectrum of $\D~^2$.

\nin First we will prove an important identity of the gamma matrices that we shall need to find the minimum polynomial of $\phi$, this identity is

\be \label{x4}
\gamma_{a_1 \cdots a_n}\gamma_b = \gamma_{a_1 \cdots a_n b} + n \gamma_{[ a_1\cdots a_{n-1} }\delta_{a_n ] b}
\ee
or its left analogue (which is proven likewise)
\be \nn
\gamma_{b}\gamma_{a_1 \cdots a_n} = \gamma_{b a_1 \cdots a_n} + n \delta_{b [ a_1} \gamma_{a_2 \cdots a_n ]}
\ee
It is easiest to prove (\ref{x4}) by looking at cases, by definition $a_1, \cdots, a_n$ are all distinct, then either $b \neq a_k$ for all $1 \leq k \leq n$ or $b = a_{r}$ for some $r$. In the first case the r.h.s gives trivially $\gamma_{a_1 \cdots a_n b}$ and the delta term vanishes, in the second case it is the first term that vanishes on the r.h.s. and to account for the second term rewrite $\gamma_{a_1 \cdots a_n} = \gamma_{a_1} \cdots \gamma_{a_r} \cdots \gamma_{a_n}$ (since all $a_k$ are distinct) and pull $\gamma_b$ through as many terms as required to reach $\gamma_{a_r}$, yielding $\pm \gamma_{a_1 \cdots \hat{a}_r \cdots a_n}$; by keeping track of the sign and normalisation one may rewrite this as (\ref{x4}).

\nin Now we shall prove an identity involving the complex structure $J_{ab}$ that we will later need, namely

\be \nn
J_{[a_1 b_1}J_{a_2 b_2} \cdots J_{a_k b_k]}J_{b_k a_k} = \Big(\frac{2(k-1)-\Tr P}{2k-1} \Big) J_{[a_1 b_1}J_{a_2 b_2} \cdots J_{a_{k-1} b_{k-1}]},~~~~~J^2=-P
\ee
the proof follows:

\be \label{x7}
J_{[a_1 b_1}J_{a_2 b_2} \cdots J_{a_k b_k]}J_{b_k a_k} =
J_{[a_1 b_1}J_{a_2 b_2} \cdots J_{a_k] b_k}J_{b_k a_k},
\ee
in view of the antisymmetry of $J$, $J^T = -J$.

\nin We rewrite this equation separating the terms that contain $\Tr P = P_{a_k a_k}$ from the ones that do not, the resulting sums are very simple since most terms are equal; equation (\ref{x7}) becomes

\beqa \nn
&-&\frac{1}{(2k-1)!}\sum_{\epsilon =1}^{(2k-1)!} \mbox{sgn}(\epsilon) J_{\epsilon_1 \epsilon_2} J_{\epsilon_3 \epsilon_4} \cdots J_{\epsilon_{2k-3} \epsilon_{2k-2} } P_{\epsilon_{2k-1} a_k} \\ \nn
&=& -\frac{1}{(2k-1)!}\Big( (2k-2)! J_{[a_1 b_1}J_{a_2 b_2} \cdots J_{a_{k-1} b_{k-1}]}P_{a_k b_k} \\ \nn
&-&J_{[a_1 b_1}\cdots J_{a_{k-1} \hat{a_k}}P_{b_{k-1}]a_k}\times(2k-2)! \\ \nn
&+&J_{[a_1 b_1}\cdots J_{\hat{a_k}a_{k-1}}P_{b_{k-1}]a_k} \times(2k-2)! \\ \nn
&-&J_{[a_1 b_1} \cdots J_{a_{k-2} \hat{a_k}}J_{a_{k-1}b_{k-1}} P_{b_{k-2}] a_k} \times(2k-2)! \\ \nn
&+&J_{[a_1 b_1} \cdots J_{\hat{a_k} a_{k-2}}J_{a_{k-1}b_{k-1}} P_{b_{k-2}] a_k} \times(2k-2)! +\cdots \Big) \\ \nn
&=&-\frac{\Tr P}{2k-1} J_{[a_1 b_1} \cdots J_{a_{k-1}b_{k-1} ]} + \frac{2(2k-2)!}{(2k-1)!} \sum_{j=1}^{k-1}J_{[a_1 b_1}\cdots J_{a_{k-1}b_{k-1}]}\\ \nn
&=&\Big(\frac{2(k-1)-\Tr P}{2k-1} \Big) J_{[a_1 b_1}J_{a_2 b_2} \cdots J_{a_{k-1} b_{k-1}]} \\ \nn
\eeqa
as promised. The symbol $\hat{~}$ over an index indicates that it is not to be affected by the antisymmetrisation bracket $[~]$.

\nin We can insert the value of the rank of $P$, $\Tr P = 2N$ for $\CP^N$ and get

\be \label{x8}
J_{[a_1 b_1}J_{a_2 b_2} \cdots J_{a_k b_k]}J_{b_k a_k} = \Big(\frac{2(k-1-N)}{2k-1} \Big) J_{[a_1 b_1}J_{a_2 b_2} \cdots J_{a_{k-1} b_{k-1}]}.
\ee
Notice that this quantity vanishes if $k \geq N+2$ since the number of independent components of $J_{ab}$ is only $2N$.

\nin By using (\ref{x4}) repeatedly one arrives at the identity

\beqa  \label{x9} 
\gamma_{a_1 a_2 \cdots a_{n}}\gamma_{b_1}\gamma_{b_2} &=& \gamma_{a_1 a_2 \cdots a_n b_1 b_2}+(n+1)\gamma_{[a_1 a_2 \cdots a_n} \delta_{b_1] b_2} \\ \nn
&+&n\gamma_{[ a_1 a_2 \cdots a_{n-1}\hat{b}_2 }\delta_{a_n ] b_1} + n(n-1) \gamma_{[a_1 a_2 \cdots a_{n-2}}\delta_{a_{n-1} \hat{b}_2} \delta_{a_n] b_1}. \\ \nn
\eeqa
Define now the $SU(N+1)$ invariants given as

\be \nn
\mathfrak{I}_k := \frac{1}{(4i)^k}J_{a_1 b_1} \cdots J_{a_k b_k} \gamma_{a_1 b_1 \cdots a_k b_k},
\ee
observe that $\mathfrak{I}_{0}=1$, $\mathfrak{I}_{1} = \phi$ and $\mathfrak{I}_{k}=0$ if $k \geq N+1$.

\nin Writing $\mathfrak{I}_{1} = \frac{1}{4i}J_{ab}\gamma_{ab} = \frac{1}{4i}J_{ab} \gamma_a \gamma_b$ and contracting (\ref{x9}) with the appropriate set of $J$'s we find that only the term with two $\delta$'s survives in the product

\be \label{x10}
\mathfrak{I}_{k}\mathfrak{I}_{1} = \mathfrak{I}_{k+1} + \frac{k}{4}(N+1-k)\mathfrak{I}_{k-1}.
\ee
We used (\ref{x8}) in obtaining this recoursive relation.
The system of equations (\ref{x10}) terminates by the properties of $\mathfrak{I}_{k}$, and it can be re-expressed all in terms of $\mathfrak{I}_{1}=\phi$ giving thus the minimum polynomial of $\phi$ that we seek.
We can now particularise to our case of interest, $\CP^2$, i.e. $N=2$:

\nin For $\CP^2$ the system (\ref{x10}) is 

\beqa \nn
\mathfrak{I}_{0} &=& \boldsymbol{1},~~~~\mathfrak{I}_{1}=\phi \\ \nn
\mathfrak{I}_{1}\mathfrak{I}_{1}&=&\mathfrak{I}_{2}+\frac{1}{2}\mathfrak{I}_{0}\\ \nn
\mathfrak{I}_{2}\mathfrak{I}_{1}&=&\frac{1}{2}\mathfrak{I}_{1}\nn
\eeqa
and gives the following minimum polynomial for $\phi$:

\be \nn
\phi (\phi^2 - 1) = 0.
\ee
This means that $\phi$ has eigenvalues $0,\pm1$ with a given degeneracy, which can be calculated directly by taking the trace of projectors onto the different eigensubspaces and compared with (\ref{D4}) 

\beqa  \nn
\Tr (1-\phi^2)&=& \mbox{deg} (0) ~=~ 2 \times (3+1) = 8 \\
\Tr \Big(\frac{\phi(\phi \pm 1)}{2}\Big)&=&\mbox{deg}(\pm1)~=~2\times(2)=4. \nn
\eeqa
It is also easy to verify from the orthogonality of the basis for matrices composed of all products of gamma matrices that the relations

\beqa \nn \label{x11}
\Tr \phi &=& \Tr (\phi \gamma) = \Tr (\phi^2 \gamma) = 0 \\ 
\Tr (\phi^2) &=& 8,~~~\mbox{with}~~~~\gamma = \prod_{a=1}^8 \gamma_{a}
\eeqa
hold.

\section{Evaluation of $S(U(2) \times U(1))$ Casimir } \label{C}

In this appendix we collect the proof of some results mentioned in sections \ref{spinspec} and \ref{orbit}.

We prove first that the quadratic Casimir operator of $H$ has the definite value on $im(\pi_-)$:

\be  \nn
C_2 (S(U(2) \times U(1)) ) \pi_- = \frac{3}{2} \pi_-
\ee

\vspace{0.2cm}

{\it Proof:} As we know $C_2 (S(U(2) \times U(1)))= T_a T_b (\delta_{ab} - P_{ab})$ is a rotationally invariant quantity, hence by equivariance it is enough to show that the result is true in the north pole since we may carry out a rotation using the elements $g \in SU(3)$ in the Clifford representation to transport our identity to any point of $\CP^2$, because

\beqa \nn
C_2 (S(U(2) \times U(1))) &=& \mathfrak{g} C^0_2 (S(U(2) \times U(1))) \mathfrak{g}^{\dagger} \\ \nn
\pi_- &=& \mathfrak{g} \pi_-^0 \mathfrak{g}^{\dagger}
\eeqa
holds for $g$ defined by (\ref{A1}) and its Clifford representation image $Cliff(g) = \mathfrak{g}$.
Computing explicitely the required generators

\beqa \nn
T_1 = \frac{1}{2i} (\gamma_{23}+ \half \gamma_{47} + \half \gamma_{65} ), & ~& T_2 = \frac{1}{2i} (\gamma_{31}+ \half \gamma_{46} + \half \gamma_{57} )  \\ \nn
T_3 = \frac{1}{2i} (\gamma_{12}+ \half \gamma_{45} + \half \gamma_{76} ), & ~& T_8 = \frac{\sqrt{3}}{4i} (\gamma_{45}+ \gamma_{67}) 
\eeqa
we find $\phi^0 = \frac{i}{2}(\gamma_{45} + \gamma_{67}) $ and 

\be \nn
C^0_2 (S(U(2) \times U(1)))  = \frac{3}{2} + \frac{1}{4}( - \gamma_{2347}+\gamma_{2356}+\gamma_{1346}+ \gamma_{1357}- \gamma_{1245}+\gamma_{1267})
\ee
a direct computation shows that $C^0_2 (S(U(2) \times U(1))) \phi^0 = \frac{3}{2} \phi^0$, this together with the fact $[ \gamma, T_a ] = 0$ completes the proof.  $\blacksquare$

In addition, the chirality satisfies
\be \nn
\phi^2 = \half (1- \Gamma)
\ee 
{\it Proof:} Following the line of reasoning we presented it is enough to show it in the north pole because from our definition
$\Gamma = \mathfrak{g} \Gamma^0 \mathfrak{g}^{\dagger}$ with $\Gamma^0 =  \gamma_{4567}$. Using the expression for $\phi^0$ above, this is trivial  $\blacksquare$.

The  Dirac operator $\D$ anticommutes with the chirality $\Gamma$: 
\be \nn
\{ \D~, \Gamma  \}=0
\ee

{\it Proof:} Since $\Gamma$ is rotationally invariant it commutes with the total ``angular momentum" $\mathcal{L}+T$, hence only need to show that it anticommutes with all four tangent gamma matrices $\gamma_a P_{ab}$, i.e. $\{ \gamma_a P_{ab}, \Gamma \}=0$.
We show it in the north pole and apply the same transport argument. This assertion is trivial in the north pole because $\gamma_{4}, \gamma_{5}, \gamma_{6}, \gamma_{7}$ obviously anticommute with $ \Gamma^0 = \gamma_{4567}$  $\blacksquare$

\newpage

In what follows we prove the assertion made in section \ref{orbit} that the only $SU(3)$ irreducible representations $(l_1, l_2)$ that contain the trivial representation when reduced under $S(U(2) \times U(1))$ are those with $l_1 = l_2$. Consider first the reduction of the fundamental representation of $SU(3)$, ${\bf 3} = {\bf 1}_{-2} \oplus {\bf 2}_{1}$, from where one sees at once that the antifundamental $\bar{\bf 3} = {\bf 1}_{2} \oplus {\bf 2}_{-1}$. In Young tableaux notation we distinguish tensor indices corresponding to the $U(1)$ charge with a $\times$ and those associated with the $SU(2)$ part of the $S(U(2) \times U(1) )$ with a $\bullet$. A bar denotes the conjugation of a diagram. Our decomposition for the ${\bf 3}, \bar{\bf 3}$ in diagrams reads:

\be \nn
\young(\ ) = \young(\cross)\oplus \young(\bull)~, \quad  \quad \overline{\young(\ )} = \overline{\young(\cross)} \oplus \overline{\young(\bull)}
\ee
In this notation it is easy to find the decomposition (\ref{D4}) for the adjoint representation $(1,1) = {\bf 8}$:

\be \label{y0}
(1,1) = \overline{\young(\ )}  \young(\ ) = \overline{\young(\bull)}  \young(\bull) \oplus \overline{\young(\cross)} \young(\bull) \oplus \overline{\young(\bull)}  \young(\cross) \oplus \overline{\young(\cross)}  \young(\cross)
\ee
It is not difficult to see that the generalization of (\ref{y0}) to an arbitrary representation $(p,q)$ containing $q$ antifundamental and $p$ fundamental representations is in fact

\be \nn
(p,q) = \bigoplus_{(j,k) = (0,0)}^{(p,q)} \overbrace{\overline{\young(\cross\Dots\cross )}}^{q-k} \! \overbrace{\overline{\young(\bull\Dots\bull )}}^{k} \! \overbrace{\young(\cross\Dots\cross)}^{p-j} \! \overbrace{\young(\bull\Dots\bull)}^{j}
\ee
From here it is straightforward to find the corresponding branching rule, if $({\bf m})_Q$ stands for the irreducible representation of $S(U(2)\times U(1))$ with dimension $m$ and $U(1)$ charge $Q$, one gets:

\be \nn
(p,q) = \bigoplus_{(j,k) = (0,0)}^{(p,q)} ({\bf k+j+1})_{2(q-p)+ 3(j-k)}
\ee
In this formula it becomes evident that only in the case $k=j=0$ and $p=q$ the trivial representation, ${\bf 1}_0$, appears in the breaking (exactly once), as claimed.

\end{appendix}

\newpage

\bibliographystyle{abbrv}

\end{document}